\catcode`@=11
\global\newcount\secno
\global\newcount\subsecno
\global\newcount\subsubsecno
\global\newcount\equationno
\global\newcount\refno
\global\newcount\footnoteno
\global\newcount\@no
\secno=0\subsecno=0\subsubsecno=0\equationno=0\refno=0\footnoteno=0
\let\\=\cr
\def\@draftleft#1{}
\def\@draftright#1{}
\overfullrule=0pt
\def\draft{\def\@draftleft##1{\leavevmode\vadjust{\smash{%
\raise3pt\llap{\eighttt\string##1~~}}}}%
\def\@draftright##1{\rlap{\eighttt~~\string##1}}%
\def\date##1{\leftline{\number\month/\number\day/\number\year\
\the\time}\bigskip}\overfullrule=5pt\normalbaselineskip=18pt
\normalbaselines}
\def\@the#1{\ifnum\the#1>0\relax\the#1\else\ifnum\the#1<0\relax
\@no=-\the#1\advance\@no'100\char\@no\fi\fi}
\def\@advance#1{\ifnum\the#1<0\global\advance#1 -1\relax
\else\global\advance#1 1\relax\fi}
\def\nsec#1\par{\bigbreak\bigskip\@advance\secno
\subsecno=0\subsubsecno=0\equationno=0
\vbox{\secfont\noindent
\@the\secno. #1\medskip}\nobreak\noindent\ignorespaces}
\def\secadvance{\@advance\secno}
\def\sec#1#2\par{\bigbreak\bigskip
\subsecno=0\subsubsecno=0\equationno=0
\if*#1\vbox{\secfont\noindent\ignorespaces#2\medskip}%
\else
\secno=#1
\vbox{\secfont\noindent\@the\secno. #2\medskip}\fi
\nobreak\noindent\ignorespaces}
\def\seclab#1{\xdef#1{\@the\secno}\@draftleft#1}
\def\nsubsec#1\par{\bigskip\@advance\subsecno
\subsubsecno=0\equationno=0
\vbox{\subsecfont\noindent\@the\secno.\@the\subsecno. #1\medskip}%
\nobreak\noindent\ignorespaces}
\def\subsecadvance{\@advance\subsecno}
\def\subsec#1#2\par{\bigskip
\subsubsecno=0\equationno=0
\if*#1\vbox{\subsecfont\noindent\ignorespaces#2\medskip}%
\else
\subsecno=#1
\vbox{\subsecfont\noindent\@the\secno.\@the\subsecno. #2\medskip}\fi
\nobreak\noindent\ignorespaces}
\def\subseclab#1{\xdef#1{\@the\secno.\@the\subsecno}%
\@draftleft#1}
\def\nsubsubsec#1\par{\medskip\@advance\subsubsecno
\vbox{\subsubsecfont\noindent
\@the\secno.\@the\subsecno.\@the\subsubsecno. #1\medskip}%
\nobreak\noindent\ignorespaces}
\def\subsubsecadvance{\@advance\subsubsecno}
\def\subsubsec#1#2\par{\medskip
\if*#1\vbox{\subsubsecfont\noindent\ignorespaces#2\medskip}%
\else
\subsubsecno=#1
\vbox{\subsubsecfont\noindent
\@the\secno.\@the\subsecno.\@the\subsubsecno. #2\medskip}\fi
\nobreak\noindent\ignorespaces}
\def\subsubseclab#1{\xdef#1{\@the\secno.\@the\subsecno.\@the\subsubsecno}%
\@draftleft#1}
\def\eqlabel#1{\@advance\equationno
\ifnum\secno=0\xdef#1{\the\equationno}\else
\ifnum\subsecno=0\xdef#1{\@the\secno.\the\equationno}\else
\xdef#1{\@the\secno.\@the\subsecno.\the\equationno}\fi\fi
\eqno({\eqnofont #1})\@draftright#1}
\def\lnlabel#1{\global\advance\equationno1
\ifnum\secno=0\xdef#1{\the\equationno}\else
\ifnum\subsecno=0\xdef#1{\@the\secno.\@the\equationno}\else
\xdef#1{\@the\secno.\@the\subsecno.\the\equationno}\fi\fi
&({\eqnofont #1})\@draftright#1}
\def\eqadvance#1{\global\advance\equationno1
\ifnum\secno=0\xdef#1{\the\equationno}\else
\ifnum\subsecno=0\xdef#1{\@the\secno.\the\equationno}\else
\xdef#1{\@the\secno.\@the\subsecno.\the\equationno}\fi\fi}
\def\eqlabelno(#1#2){\eqno({\eqnofont #1#2})\@draftright#1}
\def\lnlabelno(#1#2){&({\eqnofont #1#2})\@draftright#1}
\newwrite\rfile
\def\nref#1#2{\global\advance\refno1\xdef#1{\the\refno}%
\immediate\write
\rfile{\noexpand\item{#1.}\noexpand\@draftleft\noexpand#1%
#2}}
\def\sref#1#2{\immediate\write
\rfile{\noexpand\item{#1.}\noexpand\@draftleft\noexpand#1%
#2}}
\def\refs#1#2 {\if.#2#2$^{\rm#1}$\spacefactor=\sfcode`.{}\space
\else\if,#2#2$^{\rm#1}$\spacefactor=\sfcode`,{}\space
\else\if;#2#2$^{\rm#1}$\spacefactor=\sfcode`;{}\space
\else\if:#2#2$^{\rm#1}$\spacefactor=\sfcode`:{}\space
\else\if?#2#2$^{\rm#1}$\spacefactor=\sfcode`?{}\space
\else\if!#2#2$^{\rm#1}$\spacefactor=\sfcode`!{}\space
\else
$^{\rm#1}$#2\space\fi\fi\fi\fi\fi\fi}
\def\Refs#1{$\rm#1$}
\def\reportno#1{\line{\hfil\vbox{\halign{\strut##\hfil\cr#1\crcr}}}}
\def\Title#1{\vskip3\bigskipamount\line{\titlefont
\hfil\vbox{\halign{\strut\hfil##\hfil\cr#1\crcr}}\hfil}%
\vskip2\bigskipamount}
\def\author#1{\centerline{\authorfont#1}\medskip}
\def\address#1{\centerline{\vbox{\halign
{\strut\hfil\addressfont##\hfil\cr#1\crcr}}}%
\bigskip}
\def\abstract#1{{\narrower\abstractfont\null\bigskip\noindent\ignorespaces
#1\bigskip}}
\def\date#1{\leftline{#1}\bigskip}
\immediate\openout\rfile=refs.tmp
\font\seventeenrm=cmr17 \font\fourteenrm=cmr10 scaled 1440
\font\twelverm=cmr12  \font\eightrm=cmr8  \font\sixrm=cmr6
\font\seventeeni=cmmi10 scaled 1728 \font\fourteeni=cmmi10 scaled 1440
\font\twelvei=cmmi12  \font\eighti=cmmi8  \font\sixi=cmmi6
\font\seventeensy=cmsy10 scaled 1728 \font\fourteensy=cmsy10 scaled 1440
\font\twelvesy=cmsy10 scaled 1200 \font\eightsy=cmsy8 \font\sixsy=cmsy6
\font\seventeenbf=cmbx10 scaled 1728 \font\fourteenbf=cmbx10 scaled 1440
\font\twelvebf=cmbx12 \font\eightbf=cmbx8 \font\sixbf=cmbx6
\font\seventeentt=cmtt10 scaled 1728 \font\fourteentt=cmtt10 scaled 1440
\font\twelvett=cmtt12 \font\eighttt=cmtt8
\font\seventeenit=cmti10 scaled 1728 \font\fourteenit=cmti10 scaled 1440
\font\twelveit=cmti12 \font\eightit=cmti8
\font\seventeensl=cmsl10 scaled 1728 \font\fourteensl=cmsl10 scaled 1440
\font\twelvesl=cmsl12 \font\eightsl=cmsl8
\font\seventeenex=cmex10 scaled 1728 \font\fourteenex=cmex10 scaled 1440
\font\twelveex=cmex10 scaled 1200
\def\tenpoint{\def\rm{\fam0\tenrm}%
\textfont0=\tenrm\scriptfont0=\sevenrm\scriptscriptfont0=\fiverm
\textfont1=\teni\scriptfont1=\seveni\scriptscriptfont1=\fivei
\textfont2=\tensy\scriptfont2=\sevensy\scriptscriptfont2=\fivesy
\textfont3=\tenex\scriptfont3=\tenex\scriptscriptfont3=\tenex
\textfont\itfam=\tenit\def\it{\fam\itfam\tenit}%
\textfont\slfam=\tensl\def\sl{\fam\slfam\tensl}%
\textfont\ttfam=\tentt\def\tt{\fam\ttfam\tentt}%
\textfont\bffam=\tenbf\scriptfont\bffam=\sevenbf
\scriptscriptfont\bffam=\fivebf\def\bf{\fam\bffam\tenbf}%
\def\small{\eightpoint}%
\def\large{\twelvepoint}%
\def\normalbaselines{\lineskip\normallineskip
  \baselineskip\normalbaselineskip \lineskiplimit\normallineskiplimit}
\setbox\strutbox=\hbox{\vrule height8.5pt depth3.5pt width0pt}%
\normalbaselines\rm}
\def\twelvepoint{\def\rm{\fam0\twelverm}%
\textfont0=\twelverm\scriptfont0=\eightrm\scriptscriptfont0=\sixrm
\textfont1=\twelvei\scriptfont1=\eighti\scriptscriptfont1=\sixi
\textfont2=\twelvesy\scriptfont2=\eightsy\scriptscriptfont2=\sixsy
\textfont3=\twelveex\scriptfont3=\twelveex\scriptscriptfont3=\twelveex
\textfont\itfam=\twelveit\def\it{\fam\itfam\twelveit}%
\textfont\slfam=\twelvesl\def\sl{\fam\slfam\twelvesl}%
\textfont\ttfam=\twelvett\def\tt{\fam\ttfam\twelvett}%
\textfont\bffam=\twelvebf\scriptfont\bffam=\eightbf
\scriptscriptfont\bffam=\sixbf\def\bf{\fam\bffam\twelvebf}%
\def\small{\tenpoint}%
\def\large{\fourteenpoint}%
\def\normalbaselines{\lineskip1.2\normallineskip
  \baselineskip1.2\normalbaselineskip \lineskiplimit1.2\normallineskiplimit}
\setbox\strutbox=\hbox{\vrule height10.2pt depth4.2pt width0pt}%
\normalbaselines\rm}
\def\fourteenpoint{\def\rm{\fam0\fourteenrm}%
\textfont0=\fourteenrm\scriptfont0=\tenrm\scriptscriptfont0=\sevenrm
\textfont1=\fourteeni\scriptfont1=\teni\scriptscriptfont1=\seveni
\textfont2=\fourteensy\scriptfont2=\tensy\scriptscriptfont2=\sevensy
\textfont3=\fourteenex\scriptfont3=\fourteenex\scriptscriptfont3=\fourteenex
\textfont\itfam=\fourteenit\def\it{\fam\itfam\fourteenit}%
\textfont\slfam=\fourteensl\def\sl{\fam\slfam\fourteensl}%
\textfont\ttfam=\fourteentt\def\tt{\fam\ttfam\fourteentt}%
\textfont\bffam=\fourteenbf\scriptfont\bffam=\tenbf
\scriptscriptfont\bffam=\fivebf\def\bf{\fam\bffam\fourteenbf}%
\def\small{\twelvepoint}%
\def\large{\seventeenpoint}%
\def\normalbaselines{\lineskip1.44\normallineskip
  \baselineskip1.44\normalbaselineskip \lineskiplimit1.44\normallineskiplimit}
\setbox\strutbox=\hbox{\vrule height12.24pt depth5.04pt width0pt}%
\normalbaselines\rm}
\def\seventeenpoint{\def\rm{\fam0\seventeenrm}%
\textfont0=\seventeenrm\scriptfont0=\twelverm\scriptscriptfont0=\eightrm
\textfont1=\seventeeni\scriptfont1=\twelvei\scriptscriptfont1=\eighti
\textfont2=\seventeensy\scriptfont2=\twelvesy\scriptscriptfont2=\eightsy
\textfont3=\seventeenex\scriptfont3=\seventeenex
\scriptscriptfont3=\seventeenex
\textfont\itfam=\seventeenit\def\it{\fam\itfam\seventeenit}%
\textfont\slfam=\seventeensl\def\sl{\fam\slfam\seventeensl}%
\textfont\ttfam=\seventeentt\def\tt{\fam\ttfam\seventeentt}%
\textfont\bffam=\seventeenbf\scriptfont\bffam=\twelvebf
\scriptscriptfont\bffam=\eightbf
\def\bf{\fam\bffam\seventeenbf}%
\def\small{\fourteenpoint}%
\def\large{\seventeenpoint}%
\def\normalbaselines{\lineskip1.73\normallineskip
  \baselineskip1.73\normalbaselineskip \lineskiplimit1.73\normallineskiplimit}
\setbox\strutbox=\hbox{\vrule height14.7pt depth6.0pt width0pt}%
\normalbaselines\rm}
\def\eightpoint{\def\rm{\fam0\eightrm}%
\textfont0=\eightrm\scriptfont0=\sixrm\scriptscriptfont0=\fiverm
\textfont1=\eighti\scriptfont1=\sixi\scriptscriptfont1=\fivei
\textfont2=\eightsy\scriptfont2=\sixsy\scriptscriptfont2=\fivesy
\textfont3=\tenex\scriptfont3=\tenex\scriptscriptfont3=\tenex
\textfont\itfam=\eightit\def\it{\fam\itfam\eightit}%
\textfont\slfam=\eightsl\def\sl{\fam\slfam\eightsl}%
\textfont\ttfam=\eighttt\def\tt{\fam\ttfam\eighttt}%
\textfont\bffam=\eightbf\scriptfont\bffam=\sixbf
\scriptscriptfont\bffam=\fivebf\def\bf{\fam\bffam\eightbf}%
\def\small{\eightpoint}%
\def\large{\tenpoint}%
\def\normalbaselines{\lineskip.8\normallineskip
  \baselineskip.8\normalbaselineskip \lineskiplimit.8\normallineskiplimit}
\setbox\strutbox=\hbox{\vrule height7pt depth3pt width0pt}%
\normalbaselines\rm}
\def\small{\eightpoint}
\def\large{\twelvepoint}
\def\vfootnote#1{\insert\footins\bgroup
  \interlinepenalty\interfootnotelinepenalty
  \splittopskip\ht\strutbox 
  \splitmaxdepth\dp\strutbox \floatingpenalty\@MM
  \leftskip\z@skip \rightskip\z@skip \spaceskip\z@skip \xspaceskip\z@skip
  \footnotefont\textindent{#1}\footstrut\futurelet\next\fo@t}
\def\nfootnote{\advance\footnoteno1\@no=\footnoteno\advance\@no'140
\footnote{$^{\char\@no}$}}
\def\nvfootnote#1{\advance\footnoteno1\@no=\footnoteno\advance\@no'140
\def#1{$^{\char\@no}$}\vfootnote{$^{\char\@no}$}}
\def\titlefont{\seventeenpoint\rm}
\def\authorfont{\twelvepoint\rm}
\def\addressfont{\tenpoint\it}
\def\abstractfont{\eightpoint\rm}
\def\secfont{\tenpoint\bf}
\def\subsecfont{\tenpoint\sl}
\def\subsubsecfont{\tenpoint\it}
\def\eqnofont{\tenpoint\rm}
\def\footnotefont{\eightpoint\rm}
\catcode`@=12

\input pictex
\let\backslash=\\
\let\\=\cr
\def\e{{\rm e}}
\def\i{{\rm i}}
\def\d{\partial}

\def\ve{\varepsilon}

\def\vt{\vartheta}
\def\Re{\mathop{\rm Re}\nolimits}
\def\Im{\mathop{\rm Im}\nolimits}

\def\sh{\mathop{\rm sh}\nolimits}

\def\W(#1,#2;#3,#4|#5){\mathop W\left[\matrix{#4&#3\cr#1&#2}\bigg|
                        \matrix{#5}\right]}
\def\L(#1,#2;#3,#4|#5){\mathop L\left[\matrix{#4&#3\cr#1&#2}\bigg|
                        \matrix{#5}\right]}
\def\K(#1,#2;#3,#4|#5){\mathop K\left[\matrix{#4&#3\cr#1&#2}\bigg|
                        \matrix{#5}\right]}
\def\bul{\par\indent\llap{$\bullet$~~}\ignorespaces}
\catcode`\@=11

\def\ssubsec#1#2\par{\vskip-\lastskip \medskip\bigskip
\subsubsecno=0
\if*#1\vbox{\subsecfont\noindent\ignorespaces#2\medskip}\else
	\vbox{\subsecfont\noindent\@the\secno.#1. #2\medskip}\fi
	\nobreak\noindent\ignorespaces}
\catcode`\@=12
\def\ar#1 #2, #3 #4 {\arrow <0.3truecm> [0.1,0.3] from #1 #2 to #3 #4 }
\catcode`!=11
\def\ln#1 #2, #3 #4 {\!start(#1,#2)\!ljoin(#3,#4)}
\catcode`!=12
\def\rl#1 #2, #3 #4 {\putrule from #1 #2 to #3 #4 }
\def\trl#1 #2, #3 #4 {\linethickness .8pt
	\putrule from #1 #2 to #3 #4 \linethickness .4pt}
\def\point#1 #2 {\put{\hbox{\kern -1pt .}} [Bl] at #1 #2 }
\def\bfpoint#1 #2 {\put{\hbox{\kern -2pt \raise -2.3pt
	\hbox{$\textstyle\bullet$}}} [Bl] at #1 #2 }
\def\figcap#1#2\par{\medskip{\narrower\eightpoint\noindent
	Fig.~#1. \ignorespaces#2\par}\medskip}

\reportno{LANDAU-98-TMP-1\\hep-th/9805216}
\Title{Nearest Neighbor Two-Point Correlation Function\\
of the $Z$-Invariant Eight-Vertex Model}
\author{Michael Lashkevich$^1$ and Yaroslav Pugai$^{1,2}$}
\address{$^1$L.~D.~Landau Institute for Theoretical Physics,\\
142432 Chernogolovka, Russia \\
and\\
$^2$Department of Mathematics, University of Melbourne,\\
Parkville, Victoria 3052, Australia}
\abstract{The nearest neighbor two-point correlation function
of the $Z$-invariant inhomogeneous eight-vertex model
in the thermodynamic limit is computed using the
free field representation.
}
\date{May 1998}
\nref\Baxter{R.~J.~Baxter, {\it Exactly Solved Models in
        Statistical Mechanics}, Academic Press, 1982}
\nref\Baxcor{R.~J.~Baxter, {\it Phil.\ Trans.\ Royal Soc.\ London}\
        {\bf 289}, 315 (1978)}
\nref\LaPu{M.~Lashkevich and Ya.~Pugai, Free field construction
	for correlation functions of the eight vertex model,
	{\it Nucl.\ Phys.}\ {\bf B516}, 623 (1998)
	(\hbox{\tt hep-th/9710099})}
\nref\DFJMN{B.~Davies, O.~Foda, M.~Jimbo, T.~Miwa, and A.~Nakayashiki,
	{\it Commun.\ Math.\ Phys.}\ {\bf 151}, 89 (1993)}
\nref\JMbook{M.~Jimbo and T.~Miwa, {\it Algebraic Analysis of Solvable
	Lattice Models}, CBMS Regional Conference Series in Mathematics,
	{\bf 85}, AMS, 1994}
\nref\JMeightvertex{M.~Jimbo, T.~Miwa, and A.~Nakayashiki,
	{\it J.~Phys.}\ {\bf A26} 2199 (1993)}
\nref\LP{S.~Lukyanov and Ya.~Pugai, {\it Nucl.\ Phys.}\ {\bf B[FS]473},
        631 (1996) (\hbox{\tt hep-th/9602074})}
\nref\BK{R.~J.~Baxter and S.~B.~Kelland, {\it J.~Phys.}\ {\bf C7}, L403
	(1974)}
\nref\JMbos{M.~Jimbo, K.~Miki, T.~Miwa, and A.~Nakayashiki,
	{\it Phys.\ Lett.}\ {\bf A168}, 256 (1992)}
\nref\Lashdisorder{M.~Yu.~Lashkevich, {\it Mod.\ Phys.\ Lett.}\ {\bf B10},
       101 (1996) (\hbox{\tt hep-th/9408131})}
\nref\JMgapless{M.~Jimbo, T.~Miwa, {\it J.~Phys.}\ {\bf A29} 2923 (1996)
	(\hbox{\tt hep-th/96011135})}
\nref\Lukyanov{S.~Lukyanov, {\it Commun.\ Math.\ Phys.}\ {\bf 167},
	183 (1995) (\hbox{\tt hep-th/9307196});
	{\it Phys.\ Lett.}\ {\bf B325},
        409 (1994) (\hbox{\tt hep-th/9311189})}
\nref\BaxterEnting{R.~J.~Baxter, I.~G.~Enting {\it J.~Phys.}\ {\bf A11},
	2463 (1978)}
\nref\JMrsos{O.~Foda, M.~Jimbo, T.~Miwa, K.~Miki, and A.~Nakayashiki,
       {\it J.\ Math.\ Phys.}\ {\bf 35}, 13 (1994)}
\nref\Martinez{J.~R.~Reyes Martinez, {\it Phys.\ Lett.}\ {\bf A227}
	203 (1997) (\hbox{\tt hep-th/9609135})}
\nref\JKOStwi{M.~Jimbo, H.~Konno, S.~Odake, J.~Shiraishi,
	Quasi-Hopf twistors for elliptic quantum groups,
	\hfil{\penalty-500}\hbox{\tt q-alg/9712029} (December 1997);
	M.~Jimbo, H.~Konno, S.~Odake, J.~Shiraishi,
        Elliptic algebra $U_{q,p}(\hat{sl}_2)$:
	Drinfeld currents and vertex operators, \hbox{\tt q-alg/9802002}
	(February 1998)}

Recently, a free field construction
for correlation functions of the ($Z$-invariant) eight-vertex
model\refs{\Baxter, \Baxcor} has been proposed\refs{\LaPu} within the
algebraic approach to integrable models of statistical
mechanics\refs{\Baxter,\DFJMN-\LP}. The free field
representation provides explicit formulas for
multipoint correlation functions on the infinite lattice.
However, the resulting expressions given
in terms of a certain series
of multiple integrals turn out to be rather cumbersome.
In this letter we give an explicit expression
for the nearest neighbor two-point correlation function
in terms of a single two-fold integral, and perform some checks.
We also discuss the independence of the integral
representations of the free parameter $u_0$ of
the vertex-face correspondence entering into
the free fields construction\refs{\LaPu}.

Let us briefly recall the notations
used in Ref.~\Refs{\LaPu} (see Ref.~\Refs{\Baxter}
for a complete definition of the eight-vertex model).
The fluctuating variables
$\ve=\pm1$ are situated at edges of the square lattice.
To each configuration
of variables $\ve_1$, $\ve_2$, $\ve_3$, $\ve_4$
ordered around a vertex a local Boltzmann weight
$R_{\ve_1\ve_2}^{\ve_3\ve_4}$ is associated, as it is shown in Fig.~1a.
The nonzero Boltzmann weights can be parametrized
as follows\refs{\Baxter}
$$
\eqalign{
R(u)_{++}^{++}=R(u)_{--}^{--}=\phantom{-}a(u)
&=-\i\rho(u) \,
\textstyle
\theta_4\left(\i{\epsilon\over\pi};\i{2\epsilon r\over\pi}\right)
\theta_4\left(\i{\epsilon\over\pi}u;\i{2\epsilon r\over\pi}\right)
\theta_1\left(\i{\epsilon\over\pi}(1-u);\i{2\epsilon r\over\pi}\right),
\cr
R(u)_{+-}^{+-}=R(u)_{-+}^{-+}=\phantom{-}b(u)
&=-\i\rho(u) \,
\textstyle
\theta_4\left(\i{\epsilon\over\pi};\i{2\epsilon r\over\pi}\right)
\theta_1\left(\i{\epsilon\over\pi}u;\i{2\epsilon r\over\pi}\right)
\theta_4\left(\i{\epsilon\over\pi}(1-u);\i{2\epsilon r\over\pi}\right),
\cr
R(u)_{+-}^{-+}=R(u)_{-+}^{+-}=\phantom{-}c(u)
&=-\i\rho(u) \,
\textstyle
\theta_1\left(\i{\epsilon\over\pi};\i{2\epsilon r\over\pi}\right)
\theta_4\left(\i{\epsilon\over\pi}u;\i{2\epsilon r\over\pi}\right)
\theta_4\left(\i{\epsilon\over\pi}(1-u);\i{2\epsilon r\over\pi}\right),
\cr
R(u)_{++}^{--}=R(u)_{--}^{++}=-d(u)
&=-\i\rho(u) \,
\textstyle
\theta_1\left(\i{\epsilon\over\pi};\i{2\epsilon r\over\pi}\right)
\theta_1\left(\i{\epsilon\over\pi}u;\i{2\epsilon r\over\pi}\right)
\theta_1\left(\i{\epsilon\over\pi}(1-u);\i{2\epsilon r\over\pi}\right),
}\eqlabel\EQRmatrix
$$
where $\theta_j(u;\tau)$ is the standard
$j$th theta function with the basic periods
$1$ and $\tau$ ($\Im\tau>0$).
The normalization factor $\rho(u)$ is irrelevant for correlation
functions.

For definiteness, let us consider the
model in the antiferroelectric phase restricting
values of the parameters
$\epsilon$, $r$, $u$ to be real numbers
in the region $\epsilon>0$, $r>1$, $-1<u<1$.
For fixed $r$ the parameter $\epsilon$
measures deviation from the
criticality. In the limit $\epsilon\to 0$ the model
has a second order phase transition.
In the `low temperature' limit
$\epsilon\to\infty$ the system falls into
one of two ground states (Fig.~1b)
indexed by $i=0,1$.

Let $P_\ve^{(i)}$ be the probability in the thermodynamic
limit that the
spin in the `central' edge is fixed to be $\ve$.
The label $(i)$ indicates that spins at the edges
situated `far away' from the origin are the same as in the
$i$th ground state so that in the low-temperature limit $\epsilon\to\infty$
the probability is nonvanishing for $\ve=(-)^i$.
It has been shown in Ref.~\Refs{\LaPu} that
the one-point correlation function
$$
g_1^{(i)}=\sum_\ve \ve P_\ve^{(i)}
$$
is recovered from the bosonization procedure.
The resulting integral representation
$$
g_1^{(i)}=(-)^{i+1} 2{\vartheta'_1(0)\over\vartheta_4(0)}
\int_{C_{0}}{dv\over2\pi\i}{h_4(v)\over h_1(v)}
\eqlabel\OPcorfun
$$
can be reduced to the Baxter--Kelland formula for the
spontaneous staggered polarization\refs{\BK}.
Here we used the notations
$h_j(u)=\theta_j(u/r;\i\pi/\epsilon r)$ and
$\vartheta_j(u)=\theta_j(u;\i\pi/\epsilon)$.
The integration contour
$C_0$ goes over the imaginary period of the theta functions
(from some complex $v_0$ to $v_0+\i\pi/\epsilon$)
so that $-1<\Re v<0$.

\topinsert
%
%
\line{\hfil
\beginpicture
\setcoordinatesystem units <1cm,1cm> point at 0 -0.5
\rl 2.5 2 , 2.5 0.8 \ar 2.5 0.81 , 2.5 0.8
\rl 3 1.5 , 1.8 1.5 \ar 1.81 1.5 , 1.8 1.5
\put{$R(u-v)_{\ve_1\ve_2}^{\ve_3\ve_4}=$} [Br] at 1.2 1.4
\put{$\ve_1$} [Bl] at 2.6 0.9
\put{$\ve_2$} [Bl] at 1.9 1.7
\put{$\ve_3$} [Bl] at 2.6 1.9
\put{$\ve_4$} [Bl] at 3.0 1.6
\put{$u$} [t] at 2.5 0.6
\put{$v$} [rB] at 1.7 1.5
\put{$(a)$} [Bl]  at 1.6 -0.7
\setcoordinatesystem units <1cm,1cm> point at -4 0
\rl 1.0 3.3 , 1.0 0.7
\rl 2.0 3.3 , 2.0 0.7
\rl 3.0 3.3 , 3.0 0.7
\rl 0.7 3.0 , 3.3 3.0
\rl 0.7 2.0 , 3.3 2.0
\rl 0.7 1.0 , 3.3 1.0
\put{$+$} [lb] at 1.1 3.4
\put{$+$} [lb] at 0.4 3.1
\put{$+$} [lb] at 3.1 3.4
\put{$+$} [lb] at 2.4 3.1
\put{$+$} [lb] at 2.1 2.4
\put{$+$} [lb] at 1.4 2.1
\put{$+$} [lb] at 1.1 1.4
\put{$+$} [lb] at 0.4 1.1
\put{$+$} [lb] at 3.4 2.1
\put{$+$} [lb] at 3.1 1.4
\put{$+$} [lb] at 2.4 1.1
\put{$+$} [lb] at 2.1 0.4
\put{$-$} [lb] at 2.1 3.4
\put{$-$} [lb] at 1.4 3.1
\put{$-$} [lb] at 1.1 2.4
\put{$-$} [lb] at 0.4 2.1
\put{$-$} [lb] at 3.4 3.1
\put{$-$} [lb] at 3.1 2.4
\put{$-$} [lb] at 2.4 2.1
\put{$-$} [lb] at 2.1 1.4
\put{$-$} [lb] at 1.4 1.1
\put{$-$} [lb] at 1.1 0.4
\put{$-$} [lb] at 3.4 1.1
\put{$-$} [lb] at 3.1 0.4
\put{$(b)$} [B] at 4.0 -0.2
\setcoordinatesystem units <1cm,1cm> point at -8 0
\rl 1.0 3.3 , 1.0 0.7
\rl 2.0 3.3 , 2.0 0.7
\rl 3.0 3.3 , 3.0 0.7
\rl 0.7 3.0 , 3.3 3.0
\rl 0.7 2.0 , 3.3 2.0
\rl 0.7 1.0 , 3.3 1.0
\put{$-$} [lb] at 1.1 3.4
\put{$-$} [lb] at 0.4 3.1
\put{$-$} [lb] at 3.1 3.4
\put{$-$} [lb] at 2.4 3.1
\put{$-$} [lb] at 2.1 2.4
\put{$-$} [lb] at 1.4 2.1
\put{$-$} [lb] at 1.1 1.4
\put{$-$} [lb] at 0.4 1.1
\put{$-$} [lb] at 3.4 2.1
\put{$-$} [lb] at 3.1 1.4
\put{$-$} [lb] at 2.4 1.1
\put{$-$} [lb] at 2.1 0.4
\put{$+$} [lb] at 2.1 3.4
\put{$+$} [lb] at 1.4 3.1
\put{$+$} [lb] at 1.1 2.4
\put{$+$} [lb] at 0.4 2.1
\put{$+$} [lb] at 3.4 3.1
\put{$+$} [lb] at 3.1 2.4
\put{$+$} [lb] at 2.4 2.1
\put{$+$} [lb] at 2.1 1.4
\put{$+$} [lb] at 1.4 1.1
\put{$+$} [lb] at 1.1 0.4
\put{$+$} [lb] at 3.4 1.1
\put{$+$} [lb] at 3.1 0.4
\endpicture
\hfil}
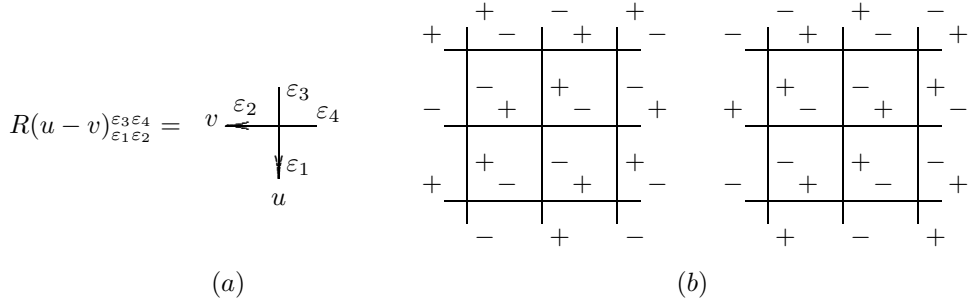
\figcap{1} $(a)$ definition of the weight matrix;
$(b)$ two degenerate ground states.\par
\endinsert

Consider now the inhomogeneous eight-vertex model on the
lattice where the spectral
parameters in two adjacent rows are
$u_1$, $u_2$ respectively, while $\epsilon$, $r$
are the same for all sites\refs{\Baxcor}.
Let $P_{\ve_1\ve_2}^{(i)}(u_1,u_2)$ denotes
the probability of the configuration with two fixed
variables as it is shown in Fig.~2a.
Here $(i)$ fixes the conditions at the
infinity so that $P_{\ve_1\ve_2}^{(i)}\to0$
unless $\ve_1=-\ve_2=(-)^i$ as $\epsilon\to\infty$.
The main statement of this letter is that
the free field construction\refs{\LaPu}
allows one to express the nearest neighbor
two-point correlation function
$$
g_2^{(i)}(u_1-u_2)\equiv\sum_{\ve_1,\ve_2=\pm}\ve_1\ve_2
P^{(i)}_{\ve_1,\ve_2}(u_1,u_2)
$$
in terms of a two-fold contour integral as follows:
$$
\eqalignno{
g_2^{(i)}(u)=
&-2{(\vartheta'_1(0))^2\over\vartheta_4(0)}
  \vt_1(u)
  {h_4(u)\over h_1(u)}
  \int_{C_{1}}
  {dv_1\over2\pi\i}
  \int_{C_{2}}
  {dv_2\over2\pi\i}
  {\vt_4(v_1+v_2-u)\over
  \vt_1(v_1-u)\vt_1(v_2-u)
}
\times
\cr
&\quad\times
    \vt_1(v_1-v_2)
    {h_4(v_1-v_2+1)\over h_1(v_1-v_2+1)}
    \prod_{j=1}^{2}{1\over\vt_1(v_j)}{h_1(v_j)\over h_4(v_j)}.
\lnlabel\TPcorfun}
$$
The integration contours $C_{1,2}$
go over the same imaginary period as
$C_0$ so that
$-1<\Re v_{1}<u<\Re v_{2}<1$,
$\Re v_{2}-\Re v_{1}<1$.

\topinsert
%
%
\line{\hfil
\beginpicture
\setcoordinatesystem units <1cm,1cm> point at 0 -0.5
\rl 0.5 3.2 , 0.5 0.1 \ar 0.5 0.1 , 0.5 0.09
\rl 1.0 3.2 , 1.0 0.1 \ar 1.0 0.1 , 1.0 0.09
\rl 1.5 3.2 , 1.5 0.1 \ar 1.5 0.1 , 1.5 0.09
\rl 2.0 3.2 , 2.0 0.1 \ar 2.0 0.1 , 2.0 0.09
\rl 2.5 3.2 , 2.5 0.1 \ar 2.5 0.1 , 2.5 0.09
\rl 3.0 3.2 , 3.0 0.1 \ar 3.0 0.1 , 3.0 0.09
\rl 3.2 0.5 , 0.1 0.5 \ar 0.1 0.5 , 0.09 0.5
\rl 3.2 1.0 , 0.1 1.0 \ar 0.1 1.0 , 0.09 1.0
\rl 3.2 1.5 , 0.1 1.5 \ar 0.1 1.5 , 0.09 1.5
\rl 3.2 2.0 , 0.1 2.0 \ar 0.1 2.0 , 0.09 2.0
\rl 3.2 2.5 , 0.1 2.5 \ar 0.1 2.5 , 0.09 2.5
\rl 3.2 3.0 , 0.1 3.0 \ar 0.1 3.0 , 0.09 3.0
\ln 1.7 1.45 , 1.8 1.55 \ln 1.8 1.45 , 1.7 1.55
\ln 1.7 1.95 , 1.8 2.05 \ln 1.8 1.95 , 1.7 2.05
\put {$\ve_1$} [cb] at 1.75 1.6
\put {$\ve_2$} [cb] at 1.75 2.1
\put {$u$} [tc] at 0.5 0.0
\put {$u$} [tc] at 1.0 0.0
\put {$u$} [tc] at 1.5 0.0
\put {$u$} [tc] at 2.0 0.0
\put {$u$} [tc] at 2.5 0.0
\put {$u$} [tc] at 3.0 0.0
\put {$v$} [rB] at 0.0 0.5
\put {$v$} [rB] at 0.0 1.0
\put {$v_1$} [rB] at 0.0 1.5
\put {$v_2$} [rB] at 0.0 2.0
\put {$v$} [rB] at 0.0 2.5
\put {$v$} [rB] at 0.0 3.0
\put {$u_j=u-v_j$} [cB] at 1.5 -0.5
\put {$(a)$} [B] at 1.75 -0.9
\setcoordinatesystem units <1cm,1cm> point at -5.5 -1.0
\rl 0.5 2.7 , 0.5 -0.4 \ar 0.5 -0.4 , 0.5 -0.41
\rl 1.0 2.7 , 1.0 -0.4 \ar 1.0 -0.4 , 1.0 -0.41
\rl 1.5 2.7 , 1.5 -0.4 \ar 1.5 -0.4 , 1.5 -0.41
\rl 2.0 2.7 , 2.0 -0.4 \ar 2.0 -0.4 , 2.0 -0.41
\rl 2.5 2.7 , 2.5 -0.4 \ar 2.5 -0.4 , 2.5 -0.41
\rl 3.0 2.7 , 3.0 -0.4 \ar 3.0 -0.4 , 3.0 -0.41
\rl 3.5 2.7 , 3.5 -0.4 \ar 3.5 -0.4 , 3.5 -0.41
\rl 3.7 0.0 , 0.1 0.0 \ar 0.1 0.0 , 0.09 0.0
\rl 3.7 0.5 , 0.1 0.5 \ar 0.1 0.5 , 0.09 0.5
\rl 3.7 1.0 , 0.1 1.0 \ar 0.1 1.0 , 0.09 1.0
\rl 3.7 1.5 , 0.1 1.5 \ar 0.1 1.5 , 0.09 1.5
\rl 3.7 2.0 , 0.1 2.0 \ar 0.1 2.0 , 0.09 2.0
\rl 3.7 2.5 , 0.1 2.5 \ar 0.1 2.5 , 0.09 2.5
\ln 1.95 1.2 , 2.05 1.3 \ln 2.05 1.2 , 1.95 1.3
\ln 1.7 1.45 , 1.8 1.55 \ln 1.8 1.45 , 1.7 1.55
\put {$\ve_1$} [Br] at 1.95 1.15
\put {$\ve_2$} [bc] at 1.75 1.6
\put {$u$} [tc] at 0.5 -0.5
\put {$u$} [tc] at 1.0 -0.5
\put {$u$} [tc] at 1.5 -0.5
\put {$v_1$} [tc] at 2.0 -0.5
\put {$u$} [tc] at 2.5 -0.5
\put {$u$} [tc] at 3.0 -0.5
\put {$u$} [tc] at 3.5 -0.5
\put {$v$} [rB] at 0.0 0.0
\put {$v$} [rB] at 0.0 0.5
\put {$v$} [rB] at 0.0 1.0
\put {$v_2$} [rB] at 0.0 1.5
\put {$v$} [rB] at 0.0 2.0
\put {$v$} [rB] at 0.0 2.5
\put {$(b)$} [B] at 1.75 -1.4
\endpicture
\hfil}
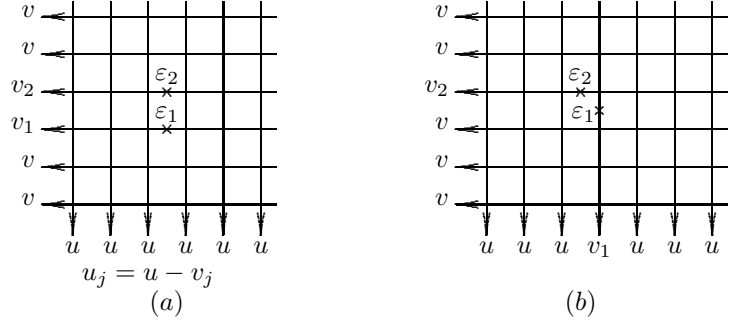
\figcap{2} Probability $P^{(i)}_{\ve_1\ve_2}(u_1,u_2)$: (a) definition;
(b) equivalent form obtained by rotating the line $v_1$.\par
\endinsert

Briefly, to get this formula
we applied the standard procedure of computing traces
of bosonic operators over Fock spaces\refs{\LP} and then
proceeded as in the one-point function
case (see Appendix~D of Ref.~\Refs{\LaPu})
by applying various identities for theta functions
to provide summation over the SOS variables including
infinite summation over Fock spaces. We will not go into
technical details of this procedure since
they are more cumbersome than instructive.

Let us only make a remark on the
free parameter $u_0$ in the free field
representation\refs{\LaPu}.
Although the explicit formulas for correlation
functions in terms of traces of bosonic fields
do contain $u_0$, it is evident from the physical grounds that
correlation functions must be $u_0$-independent.
Since the free field representation
is based on some assumptions, it
is important to prove this statement
for the resulting integral
representations for
correlation functions.
It can be easily checked that these expressions
are nonvanishing double periodic
meromorphic functions of $u_0$ with two
generically incommensurate real periods $2$ and $2r$, which proves
their $u_0$-independence. We used
the fact that $g_2^{(i)}$ is independent of $u_0$ to fix
$u_0=u_2+{\i\pi\over2\epsilon}$ in Eq.~(\TPcorfun).
In addition, to make
sure, we also obtained the formula for $g_2^{(i)}$
with general $u_0$ (which turns out to
be more cumbersome) and checked
numerically that it does not depend on $u_0$.

To support the validity of the integral
representation (\TPcorfun) we compared it
with the known results.

\bul The partition function differentiation method.\refs{\Baxcor}%
\nfootnote{We would like to thank Prof.\ J.~H.~H.~Perk for
pointing to us the possibility of this check.}
The probabilities $P^{(i)}_{\ve_1\ve_2}(u_1,u_2)$ are
equal to those shown in Fig.~2b because of the $Z$-invariance.
So the correlation function $g_2^{(i)}(u)$ can be calculated as
$$
g_2^{(i)}(u)=\left.\left(a{\d\over\d a}-b{\d\over\d b}
-c{\d\over\d c}+d{\d\over\d d}\right)\log\kappa(a,b,c,d)
\right|_{u,\epsilon,r}.
\eqlabel\TPdiff
$$
Here $\kappa$ is the partition function per site as a function
of the Boltzmann weights\refs{\Baxter},
$$
\log\kappa(a,b,c,d)=\log(c+d)
+\sum_{m=1}^\infty
{(x^{3m}-p^{m/2})(1-p^{m/2}x^{-m})(x^m+x^{-m}-z^m-z^{-m})
\over m(1-p^m)(1+x^{2m})}
$$
with $x=\e^{-\epsilon}$, $p=x^{2r}$, $z=\e^{\epsilon(1-2u)}$,
and the derivatives are taken at the point
characterized by the parameters $u$, $\epsilon$, $r$ according to
Eq.~(\EQRmatrix).

We checked numerically that the results of (\TPcorfun) and (\TPdiff)
coincide at least up to the fifth decimal digit in a wide region
of values of $\epsilon$ and $r$.


\bul The limiting case $r\to\infty$. In this limit the
Boltzmann weights (\EQRmatrix) become those of the six-vertex
model in the antiferroelectric
regime. The integral representation for correlation functions of
the antiferroelectric six-vertex model is known\refs{\JMbos,\JMbook}.
Analytically the formula (\TPcorfun) gives another integral representation
in this limit, but numerically the results of
integration according to both formulas coincide up to the sixth decimal digit.

\bul The limiting case $\epsilon\to0$.
This is the critical region of the
eight-vertex model.
Using the Baxter duality transformation
for the Boltzmann weights\refs{\Baxter}
one can map the model in this region
onto the six-vertex model in the gapless regime
whose correlation functions were
found in Ref.~\Refs{\JMgapless}
(see also Ref.~\Refs{\Lukyanov}).
Performing the duality transform at the level
of correlation functions\refs{\Lashdisorder} one has to
compare our answer with
the following correlation function
$$
g_{2}^{JM}(\beta_1-\beta_2)\equiv
-2<E^{(1)}_{-+}E^{(2)}_{+-}>(\beta_1,
\beta_2)
$$
in the notations of Ref.~\Refs{\JMgapless}
with identification $\nu=1/r$, $\beta_j= \i \pi u_j$.
The integral representation for this quantity
can be written as follows\refs{\JMgapless}
$$
\eqalignno{
g^{JM}_2(\beta)=
&-2
  {\sh\beta\over\sh\nu\beta}
  \int_{C_{-}}
  {dv_1\over2\pi\i}
  \int_{C_{+}}
  {dv_2\over2\pi\i}
  {1\over
  \sh(v_1-\beta)\sh(v_2-\beta)
}
\times
\cr
&\quad\times
    {\sh(v_1-v_2)\over \sh\nu(v_1-v_2+\i\pi)}
    \prod_{j=1}^2{\sh\nu v_j\over \sh v_j}.
\lnlabel\JMcorfun}
$$
Here the contours $C_{\pm}$ are from $-\infty$ to $+\infty$.
They are chosen in such a way that $\beta+\pi \i$
(resp.\ $\beta$) is above (resp.\ below) $C_+$ and
$\beta$ (resp.\ $\beta-\pi \i$) is above (resp.\ below) $C_-$.
By checking it directly one shows that
the
limit $\epsilon \to 0$ of Eq.~(\TPcorfun)
coincides with Eq.~(\JMcorfun).

\bul The $r=2$ case.
Under such specification the eight-vertex
model is equivalent to a two of
non-interacting  Ising
models\refs{\Baxter}. In this case $-g_2^{(i)}$ coincides
with the nearest neighbor diagonal correlator of the
inhomogeneous ($Z$-invariant) Ising
model in the ferromagnetic regime\refs{\BaxterEnting}
(see also Refs.~\Refs{\Baxcor,\JMrsos,\Martinez}),
$$
-g_2^{(i)}(u)=\langle\sigma_{m,n}\sigma_{m+1,n+1}\rangle
={1\over\pi}
{\theta_4(0;\i{2\epsilon\over\pi})\over\theta_3(0;\i{2\epsilon\over\pi})}
{\theta'_2(\i{\epsilon\over\pi}u;\i{2\epsilon\over\pi})\over
\theta_1(\i{\epsilon\over\pi}u;\i{2\epsilon\over\pi})},
\eqlabel\TPIsing
$$
where $\sigma_{m,n}$ is the spin variable at the site $(m,n)$
of the square lattice.
As we show in the Appendix, Eq.~(\TPcorfun) reduces to
this formula at $r=2$.

We hope that a similar integral representation can
be obtained starting from the free field representation
for other multipoint correlation functions, in particular,
for two-point functions with separation of 2 lattice sites or more.
It would be also very interesting to understand whether
the elliptic algebra approach proposed in Refs.~\Refs{\JKOStwi}
would lead to another bosonization prescription and
give a direct procedure of obtaining the integral representations of the
correlation functions of the eight-vertex model.

\sec* Acknowledgments

We are very grateful to M.~Jimbo and T.~Miwa for
their kind hospitality and support in Kyoto University
and RIMS where a part of this work has been
done. We would like to thank
M.~Jimbo, K.~Hasegawa, H.~Konno, T.~Miwa, S.~Odake, J.~Shiraishi
and all participants of the seminar
``Elliptic algebras and bosonization" in Kyoto for useful discussions.
We are indebted to R.~J.~Baxter, B.~Davies, O.~Foda,
S.~Lukyanov, and J.~H.~H.~Perk
for their interest to the work and valuable remarks improving the manuscript.
Ya.~P. is very grateful to R.J.~Baxter for his kind hospitality in ANU and
to O.~Foda for constant attention and support.
The work was supported, in part, by CRDF under the grant RP1--277
and by INTAS and RFBR under the grant INTAS--RFBR--95--0690,
and by RFBR under the grants 96--15--96821 and 96--02--16507.
Ya.~P.\ was also supported by the Australian Research
Council.

\secno=-1
\sec* Appendix

Let us obtain (\TPIsing) from (\TPcorfun) in the
Ising model case $r=2$.
The integrand of (\TPcorfun) is antisymmetric with respect
to the permutation of $v_1$ and $v_2$. This allows
one to perform the second integration simply
by taking the
residues at the pole $v_2=u$.
Using the identity
$$
{\vartheta_4(u)\over\vartheta_1(u)}{h_1(u)\over h_4(u)}
={\vartheta_4(0)\over\vartheta'_1(0)}{h'_1(0)\over h_4(0)}
{h_1(1)\over h_4(1)}{h_4(u+1)\over h_1(u+1)}
\qquad(\hbox{for }r=2),
$$
which is valid for $r=2$, the resulting expression
can be represented in the following form
$$
g_2^{(i)}(u)=2{h'_1(0)\over h_4(0)}{h_1(1)\over h_4(1)}J(u),
\qquad J(u)=\int_{C_1}{dv\over2\pi\i}
{h_4(v+1)h_4(v-u+1)\over h_1(v+1)h_1(v-u+1)}.
\eqlabel\Jinteg
$$
The function $J(u)$ satisfies the following
determining properties
$$
\eqalign{
\span
J\left(u+{\i\pi\over\epsilon}\right)=J(u),
\qquad J(-u)=J(u),
\qquad J\left(\i\pi\over2\epsilon\right)={1\over2\epsilon},
\cr
\span
J(u+2)=-J(u)+{h_4(u)h_4(0)\over h_1(u)h_1(0)},
\cr
\span
J(u)\hbox{ is regular on the strip }-2<\Re u<2,
}
$$
which fix it completely to be
\par\vbox{
$$
J(u)=-{1\over2\epsilon}
{\theta_2(0,\i{2\epsilon\over\pi})\over
\theta'_1(0,\i{2\epsilon\over\pi})}
{\theta'_2(\i{\epsilon\over\pi}u;\i{2\epsilon\over\pi})\over
\theta_1(\i{\epsilon\over\pi}u;\i{2\epsilon\over\pi})}.
$$
Passing to the conjugate module in the coefficient at $J(u)$ in Eq.~(\Jinteg)
one gets (\TPIsing).
}

\bigskip\allowbreak\bigskip\immediate\closeout\rfile
\vbox{\secfont\noindent References\bigskip}\nobreak
\catcode`@=11\input refs.tmp\catcode`@=12\bigskip

\end